# Investigation of the Physical Mechanism behind Retention Loss in FeFETs with MIFIFIS Gate Structure

Tao Hu, Zeqi Chen, Runhao Han, Xinpei Jia, Jia Yang, Mingkai Bai, Ruoyao Ji, Yajing Ding, Mengwei Zhao, Yuhan Li, Kaiyi Li, Wenbo Fan, Xianzhou Shao, Xiaoqing Sun, Kai Han, Jing Zhang, Yanrong Wang, Junshuai Chai, Hao Xu, Xiaolei Wang, Wenwu Wang and Tianchun Ye

*Abstract*—A Metal-Gate Blocking Layer (GBL)-Ferroelectric-Tunnel Dielectric Layer (TDL)-Ferroelectric -Channel Insulator (Ch.IL)-Si (MIFIFIS) structure is proposed to achieve a larger MW for applications in Fe-NAND. However, the large retention loss (RL) in the MIFIFIS structure restricts its application. In this work, we vary the physical thickness of the GBL and TDL, and conduct an in-depth analysis of the energy bands of the gate structure to investigate the physical mechanism behind the RL in FeFETs with the MIFIFIS structure. The physical origin of the RL is that the electric field direction across the TDL reduces the potential barrier provided by the ferroelectric near the silicon substrate. Based on the above physical mechanism, the RL can be reduced to 12% and 0.2% by redesigning the gate structure or reducing the pulse amplitude, respectively. Our work contributes to a deeper understanding of the physical mechanism behind the RL in FeFETs with the MIFIFIS gate structure. It provides guidance for enhancing the reliability of FeFETs.

## I. INTRODUCTION

SINCE the discovery of ferroelectricity in doped $HfO_2$ in 2011 [1], hafnia-based silicon channel ferroelectric field-effect transistors ($HfO_2$ Si-FeFETs) have attracted widespread research interest as a strong candidate for

This work was supported in part by the National Natural Science Foundation of China under Grant Nos. 92264104 and 52350195, in part by National Key Research and Development Program of China under Grant No. 2022YFB4400300, and Supported by the Postdoctoral Fellowship Program of CPSF under Grant No. GZC20232925. *(Tao Hu and Zeqi Chen contributed equally to this work.) (Corresponding authors: Hao Xu, Junshuai Chai.)*

Tao Hu, Zeqi Chen, Runhao Han, Xinpei Jia, Jia Yang, Mingkai Bai, Ruoyao Ji, Yajing Ding, Mengwei Zhao, Yuhan Li, Kaiyi Li, Wenbo Fan, Xianzhou Shao, Xiaoqing Sun, Junshuai Chai, Hao Xu, Xiaolei Wang, Wenwu Wang and Tianchun Ye are with the Institute of Microelectronics, Chinese Academy of Sciences, Beijing 100029, China, and also with the School of Integrated Circuits, University of Chinese Academy of Sciences, Beijing 100049, China (e-mail: xuhao@ime.ac.cn; chaijunshuai@ime.ac.cn).

Kai Han is with the School of Physics and Electronic Information, Weifang University, Weifang 261061, Chian.

Yanrong Wang and Jing Zhang are with the School of Information Science and Technology, North China University of Technology, Beijing 100144, China.

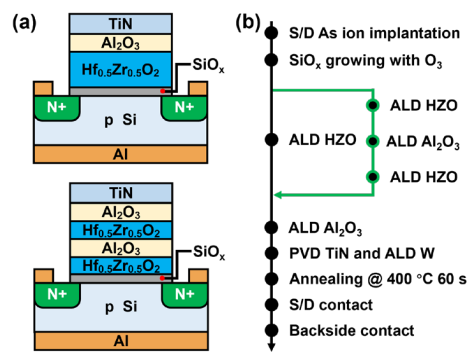

Fig. 1. (a) Schematic and (b) fabrication process flow of FeFETs.

nonvolatile memory with low write voltage, high switching speed, CMOS compatibility, and excellent scalability [2-17]. For FeFETs with the MFIS gate structure, the large amount of charge injected from the silicon channel shields the spontaneous polarization of the ferroelectric [18-21]. This results in a narrow memory window (MW), and the MW is generally limited to less than 2 V [22-25], which severely restricts its application in multi-bit memory.

Recently, a Metal-Gate Blocking Layer (GBL)-Ferroelectric -Tunnel Dielectric Layer (TDL)-Ferroelectric-Channel Insulator (Ch.IL)-Si (MIFIFIS) structure has been proposed to achieve a larger MW [26-28]. Unfortunately, the retention characteristics of the MIFIFIS gate structure exhibit a significant degradation. The retention loss (RL) exceeds 25%, which limits its application in Fe-NAND [29, 30]. Therefore, we conducted a detailed study on the physical mechanism behind the RL in the MIFIFIS gate structure. This mechanism is attributed to the direction of the electric field across TDL pointing towards the silicon substrate, which reduces the potential barrier provided by the ferroelectric near the silicon substrate.

In addition, to reverse the electric field direction across the TDL (specifically, directing it toward the silicon substrate), we propose the following strategies: (i) inserting a charge-trapping layer (CTL) on both sides of the TDL to increase the charge density at the TDL ends, or (ii) reducing the operating voltage amplitude to decrease the charge density at the GBL/FE interface. Using these methods, the RL can be reduced to 12% and 0.2%, respectively. This study provides guidance for improving FeFET reliability through gate stack engineering

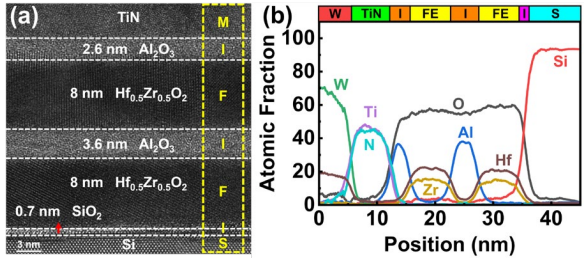

Fig. 2. (a) HRTEM images and (b) EDS of the 8483 MIFIFIS structure.

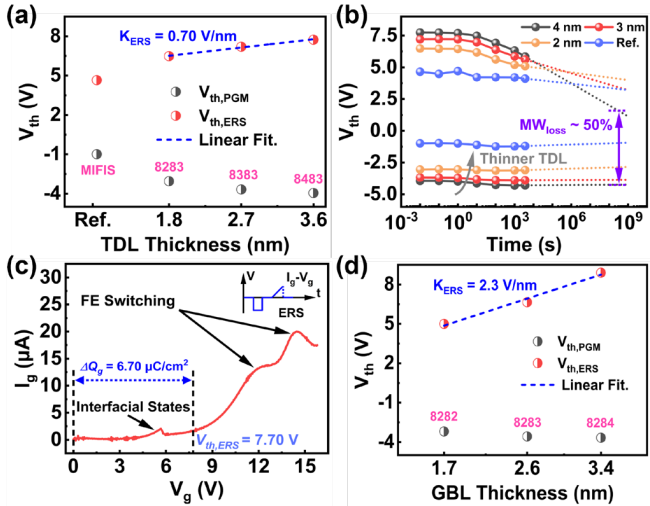

Fig. 3. The measurement results of (a) $V_{th}$ and (b) retention characteristics corresponding to the maximum MW. (c) The $I_g$-$V_g$ curve of the reading process after the erase pulse. (d) The dependence of $V_{th}$ corresponding to the maximum MW on the physical thickness of the GBL.

and optimizing pulse conditions.

## II. DEVICE FABRICATION AND CHARACTERIZATION

We fabricated FeFETs with different gate structures, as shown in Fig. 1(a). One is a TiN/Al$_2$O$_3$/Hf$_{0.5}$Zr$_{0.5}$O$_2$/SiO$_x$/Si (MIFIS) structure with 18 nm Hf$_{0.5}$Zr$_{0.5}$O$_2$ and 3 nm Al$_2$O$_3$ as the control sample, and the other is a TiN/Al$_2$O$_3$/Hf$_{0.5}$Zr$_{0.5}$O$_2$/Al$_2$O$_3$/Hf$_{0.5}$Zr$_{0.5}$O$_2$/SiO$_x$/Si (MIFIFIS) structure with 8 nm Hf$_{0.5}$Zr$_{0.5}$O$_2$. For the MIFIFIS structure, when the TDL thickness is 2 nm, the GBL thickness is 2-, 3-, or 4 nm, and when the GBL thickness is 3 nm, the TDL thickness is 2-, 3-, or 4 nm. For simplicity, we denote the gate stacks by their ferroelectric/TDL/ferroelectric/GBL nominal thicknesses. For example, 8/2/8/2 (8282), 8/2/8/3 (8283), and 8/2/8/4 (8284). Fig. 1(b) shows the fabrication process flow. This detailed process flow is described in our previous works [31, 32]. All of the fabrication processes of the MIFIS structure are the same as the MIFIFIS structure, except that the gate structure and thickness of the dielectric layer are different.

Fig. 2 shows the High-Resolution Transmission Electron Microscopy (HRTEM) images and the Energy Dispersion Spectrometer (EDS) of the 8483 MIFIFIS structure.

In this work, the gate length/width (*L/W*) of the FeFETs is 5/150 μm. The electrical measurements were performed by Keysight B1500A. The threshold voltage ($V_{th}$) is extracted by the constant current method.

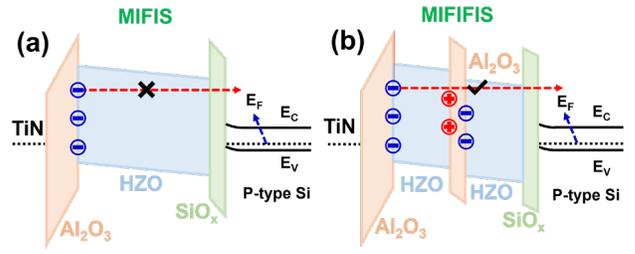

Fig. 4. The energy band diagram of (a) the MIFIS structure and (b) the 8283 MIFIFIS structure after the erase operation.

## III. RESULTS AND DISCUSSION

We first study the maximum MW of the FeFETs with different gate structures. Here, we find the maximum MW by adopting different program/erase (P/E) voltage amplitudes under the pulse width of 50 μs for each sample. Fig. 3(a) shows the dependence of $V_{th}$, which corresponds to the maximum MW, on the TDL thickness. For the MIFIS structure and the 8283 MIFIFIS structure, at the same physical thickness of the gate stack, the 8283 MIFIFIS structure shows a maximum MW of 9.5 V, while the MIFIS gate structure shows a maximum MW of 5.7 V. Therefore, inserting an Al$_2$O$_3$ interlayer in the middle of the ferroelectric Hf$_{0.5}$Zr$_{0.5}$O$_2$ can increase the MW of FeFETs, which is consistent with the phenomenon reported in [33, 34]. This is attributed to the presence of the trapped charges at both ends of the TDL [7, 33]. Furthermore, we find that $V_{th,ERS}$ increases linearly with the increase of the TDL thickness. According to (1), this slope $K_{ERS}$ in Fig. 3(a) represents the magnitude of the electric field across the TDL at $V_g$=$V_{th,ERS}$.

$$V_{th} = \varphi_{MS} + 2\varphi_B + \frac{1}{C_{TDL}} \cdot (Q_{TDL} + Q_{t\_Ch} + Q_{Si\_th}) + \frac{1}{C_{GBL}} \cdot (Q_{t\_GBL} + Q_{t\_Ch} + Q_{Si\_th}) + \frac{2}{C_{FE}} \cdot (P_{FE\_th} + Q_{t\_Ch} + Q_{Si\_th}) + \frac{1}{C_{BIL}} \cdot Q_{Si\_th} \quad (1)$$

Here, $Q_{t\_Ch}$, $Q_{t\_TDL}$, $Q_{t\_GBL}$, and $Q_{Si\_th}$ respectively represent the charge densities at the GBL/FE, TDL/FE, FE/Ch.IL interfaces, and on the silicon substrate when $V_g$=$V_{th}$.

We investigate the retention characteristics under the maximum MW. Fig. 3(b) shows the measurement results of retention characteristics. We find that the retention characteristics of the MIFIFIS structure have significantly degraded compared to the MIFIS structure, and the RL increases with the increase of the TDL thickness. In addition, we find that the RL is mainly caused by the rapid degradation of the $V_{th,ERS}$. Therefore, we need to investigate the physical origin of the $V_{th,ERS}$ degradation when $V_g$=0 V after the erase operation. The latest research results indicate that the de-trapping of the trapped charges at the GBL/FE onto the silicon substrate leads to the RL of the $V_{th,ERS}$ in the MIFIS gate structure [35, 36]. Thus, we investigate the effect of the TDL insertion and increased TDL thickness on the de-trapping potential barrier of the trapped charges.

We discuss in detail the effect of the TDL insertion on the de-trapping potential barrier through an in-depth analysis of the energy bands. We measure the same $I_g$-$V_g$ curve as the reading process after the erase operation. Fig. 3(c) shows the measurement results. We find that there is almost no switching of the ferroelectric polarization during the 0-$V_{th,ERS}$. Therefore, $I_g$ is a current formed by the background capacitance response of the gate stack. We can integrate $I_g$ to obtain $\Delta Q_g$, and the change in the electric field of each layer in the gate stack can be calculated by (2)

$$\Delta E = \frac{\Delta Q_g}{\varepsilon_0 \varepsilon_r} \quad (2)$$

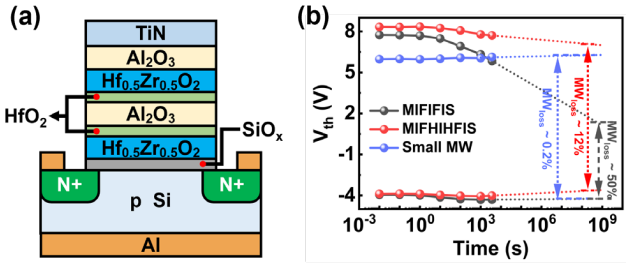

Fig. 5. (a) Schematic of the MIFHIHFIS-FeFETs. (b) Comparison between the retention characteristics.

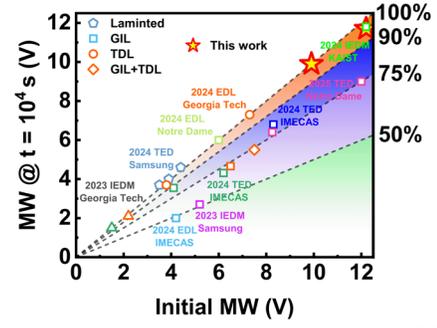

Fig. 6. The benchmark for the retention characteristics of FeFET.

Therefore, we can calculate the electric field across the TDL at $V_g$=0 V from the electric field across the TDL at $V_g$=$V_{th,ERS}$. We find that the electric field across the TDL at $V_g$=0 V is -1.41 MV/cm (specifically, directing it toward the metal gate), which is opposite to the direction of the electric field formed by the trapped charges on both sides of the TDL, and the same direction as the electric field formed by the trapped charges at the GBL/FE interface, as shown in Fig. 4. This is because $Q_{t\_GBL}$ is significantly larger than $Q_{t\_TDL}$, which is consistent with the slope in Fig. 3(d) being greater than the slope in Fig. 3(a). The electric field across the TDL at $V_g$=0 V reduces the potential barrier provided by the ferroelectric near the silicon substrate, as shown in Fig. 4, resulting in a significant RL in the MIFIFIS structure compared to the MIFIS structure.

To reverse the electric field direction across the TDL (specifically, directing it toward the silicon substrate), we propose the following strategies: (i) increasing the charge density at the TDL ends. Therefore, we insert a 1.5 nm $HfO_2$ CTL on both sides of the TDL to increase the charge density at the TDL ends. Fig. 5(a) shows the optimized device structure, namely the MIFHIHFIS structure. (ii) Reducing the charge density at the GBL/FE interface. Therefore, we reduce the erase pulse amplitude to decrease the charge density at the GBL/FE interface. Using these methods, the RL can be reduced to 12% and 0.2%, respectively, as shown in Fig. 5(b).

Fig. 6 shows the benchmark of our work. Our device demonstrates the ability to achieve both high MW and good retention characteristics simultaneously.

## IV. CONCLUSION

This work investigates the physical mechanism behind the retention loss of the MIFIFIS-FeFET. This is attributed to the direction of the electric field pointing towards the metal gate across the TDL at Vg=0 V, which reduces the potential barrier contributed by the ferroelectric near the silicon substrate, resulting in a significant RL. By adopting the following strategies: (i) inserting a CTL on both sides of the TDL and (ii) reducing the erase pulse amplitude, RL can be reduced to 12% and 0.2%, respectively. Our work contributes to a deeper understanding of the physical mechanism behind the RL in FeFETs with the MIFIFIS gate structure.